\documentclass[%
 letterpaper,
 twocolumn,
 superscriptaddress,
 amsmath,
 amssymb,
 nobibnotes,
 pra,
]{revtex4-2}

\usepackage{graphicx}
\usepackage{dcolumn}
\usepackage{bm}

\usepackage{xcolor}
\usepackage{CJK}
\usepackage{pifont}
\usepackage{cleveref}
\usepackage{verbatim}
\usepackage{comment}
\usepackage {upgreek}

\begin{document}
\begin{CJK*}{UTF8}{gbsn} 

\preprint{APS/123-QED}

\title{
Coherent Molecular Deceleration via Vibrational Bichromatic Force
}

\author{Meng-Yi Yu}%
\affiliation{Hefei National Research Center for Physical Sciences at the Microscale, University of Science and Technology of China, Hefei 230026, China}

\author{Ya-Nan Lv}
 \affiliation{Hefei National Research Center for Physical Sciences at the Microscale, University of Science and Technology of China, Hefei 230026, China}
 \affiliation{Hefei National Laboratory, University of Science and Technology of China, Hefei 230088, China}


\author{Cun-Feng Cheng}%
 \email{cfcheng@ustc.edu.cn}
 \affiliation{Hefei National Research Center for Physical Sciences at the Microscale, University of Science and Technology of China, Hefei 230026, China}
 \affiliation{Hefei National Laboratory, University of Science and Technology of China, Hefei 230088, China}

\author{Shui-Ming Hu}
 \affiliation{Hefei National Research Center for Physical Sciences at the Microscale, University of Science and Technology of China, Hefei 230026, China}
 \affiliation{Hefei National Laboratory, University of Science and Technology of China, Hefei 230088, China}


\date{\today}

\begin{abstract}
We propose a scheme for direct laser deceleration of molecules based on a vibrational transition-mediated bichromatic force (VBCF). By precisely engineering mid-infrared optical fields, we establish coherent absorption-stimulated emission cycles while exploiting the long lifetime of vibrational excited states to suppress spontaneous decay and decoherence, rendering the deceleration process effectively non-dissipative. Unlike schemes based on electronic transitions, our approach completely circumvents the restrictive Franck-Condon factors. Using the fundamental vibrational transition of $^{13}$CO$_2$ as a test case, we achieve a deceleration of $1.45\times 10^5$~m/s$^2$ with negligible population loss over the full interaction time. This VBCF framework provides a general route to cold molecules applicable to any species with an allowed fundamental vibrational transition, opening broad prospects in cold chemistry and quantum metrology.

\end{abstract}

\maketitle
\section{Introduction}

Cold molecules enable a rapidly growing range of applications~\cite{Langen2024NP}, including quantum information processing and simulation~\cite{Li2023Nature, Cornish2024NP}, precision metrology and fundamental physics~\cite{Roussy2023Science, DeMille2024NP}, and cold chemistry~\cite{Toscano2020PCCP, Markus2020PRL, Liu2022AnnualReviews, Merkt2025PRL, Osterwalder2017, Ospelkaus2010Science, Ni2010Nature, Yang2019Science}. Cyclic-transition laser cooling is the most efficient method for cooling atoms~\cite{Phillips1998RMP, Wieman1999RMP, Bloch2008RMP, Chin2010RMP}, but extending it to general molecular systems has proven challenging. Unlike atoms, molecules possess vibrational and rotational degrees of freedom that inevitably introduce dissipation into electronic transitions, making truly closed cycling transitions exceptionally rare. To date, experiments have been confined to molecules with quasi-atomic level structures possessing highly diagonal Franck-Condon factors (FCFs)~\cite{Vilas2022Nature, Collopy2018PRL, Shuman2010Nature, Lim2018PRL, Carson2022NJP, Kozyryev2017PRL, Mitra2020Science, Stuhl2008PRL, Hummon2011PRL}; such molecules support near-closed cycling transitions that mitigate population leakage into dark rovibrational states — leakage that would otherwise terminate the continuous photon scattering cycle essential for sustained deceleration and cooling.

A fundamentally different approach is to exploit molecular vibrational transitions. Unlike electronic transitions, which involve electron rearrangement, molecular vibrations originate from nuclear motion and therefore exhibit spontaneous relaxation rates orders of magnitude slower. This long excited-state lifetime lifts the Franck-Condon constraint, allowing essentially lossless optical cycling. However, the correspondingly weak radiative scattering force makes vibrational transitions impractical for conventional laser cooling --- a long-standing bottleneck.

Here, we introduce a molecular deceleration scheme that overcomes this bottleneck by combining the lossless nature of vibrational transitions with the bichromatic force (BCF). The BCF, in which two counter-propagating standing waves detuned from resonance by $\pm\delta$ drive repeated stimulated absorption and emission cycles, produces a directional force proportional to $\delta$  and can deliver momentum transfer far exceeding that of the spontaneous emission force~\cite{voitsekhovich1989observation, Grimm1994, Grimm1997PRL, Metcalf2004}. Despite pioneering efforts to apply the BCF to molecules using electronic transitions~\cite{Kozyryev2018PRL, Eyler2018PRA}, complex electronic structures and Franck-Condon constraints cause irreversible population loss, limiting those demonstrations to proof-of-principle.
Our vibrational transition-mediated bichromatic force (VBCF) offers three key advantages: (i) It exploits the intrinsically low-loss, long-lived vibrational excited states to avoid decoherence and population leakage, making the deceleration process effectively non-dissipative. (ii) It uses coherent stimulated absorption-emission cycles --- rather than spontaneous scattering --- to overcome the weak radiative force of vibrational transitions, achieving high momentum transfer rate. (iii) It is universally applicable to any molecular species with an allowed vibrational transition, regardless of Franck-Condon factors.

We demonstrate the VBCF using the fundamental vibrational transition of $^{13}$CO$_2$ as a test case. With optimal mid-infrared field engineering, we expect to achieve a deceleration of $1.45\times10^5$~m/s$^2$ and show that the scheme remains robust over a broad velocity capture range. Monte Carlo simulations confirm efficient compression of the molecular velocity distribution. This work establishes a universal, all-optical route to cold molecules, opening broad prospects in cold chemistry, quantum metrology, and molecular quantum science.


\begin{figure*}[htbp]
\centering
\includegraphics[width=0.9\linewidth]{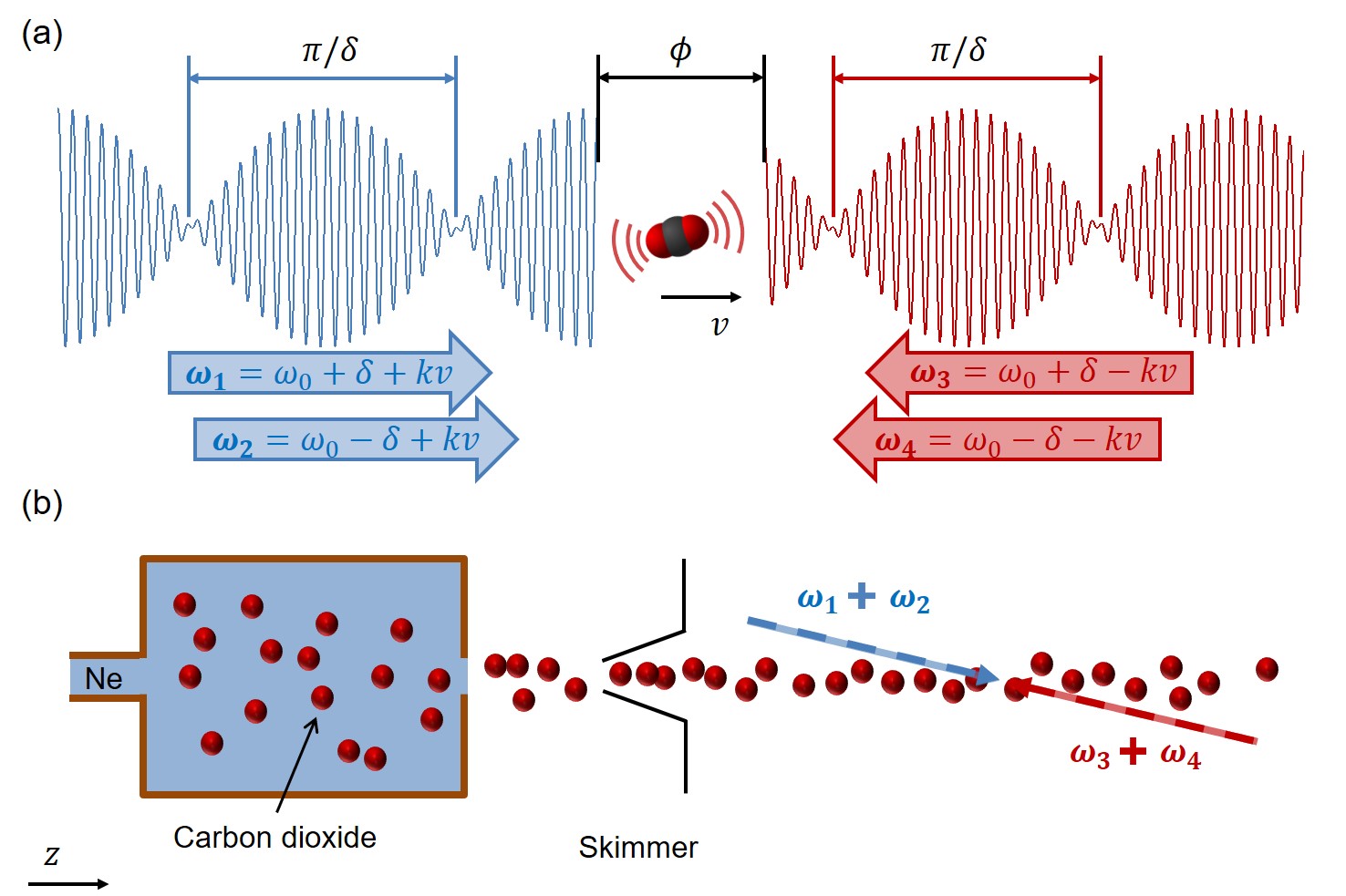}
\caption{\label{Fig:setup}
(a) The schematic diagram of the interaction between beat chains and molecules. For the molecules moving at a speed of $v$, the two counter-propagation beat note pulse trains have the detuning values of $+kv$ and $-kv$, respectively. Each pulse train is composed of two laser beams with the detuning values of $+\delta$ and $-\delta$. The period of the pulse train and the phase difference between the two pulse trains are marked in the figure. 
(b) Experimental schematic of the molecular deceleration with VBCF.
}
\end{figure*}
\section{Methodology}
Figure~\ref{Fig:setup}(a) illustrates the VBCF scheme: two counter-propagating bichromatic beat-note trains drive repeated stimulated absorption and emission cycles on a vibrational transition, transferring a net momentum of 2$\hbar k$ per cycle. For a molecule with longitudinal velocity $v$, the counter-propagating fields are Doppler-shifted by $\pm k v$ from resonance. As a demonstration, we consider $^{13}$CO$_2$ (Figure~\ref{Fig:setup}(b)) and use the R(0) (00011)-(00001) transition at $\lambda$ = 4377~nm which has a pronounced transition dipole moment. A buffer-gas-cooled beam at $150\pm 75$~m/s is collimated by a skimmer and enters the interaction region, where two phase-locked bichromatic laser beams intersect the molecular beam at a shallow angle of $1^{\circ}$, defining a 28.6~mm interaction length. Because the spontaneous emission lifetime of the vibrational excited state ($\sim$8∼ms) far exceeds the transit time ($\sim$200~$\mu$s), coherent cycling proceeds uninterrupted throughout the entire interaction.

\begin{figure*}[htbp]
\centering
\includegraphics[width=0.9\textwidth]{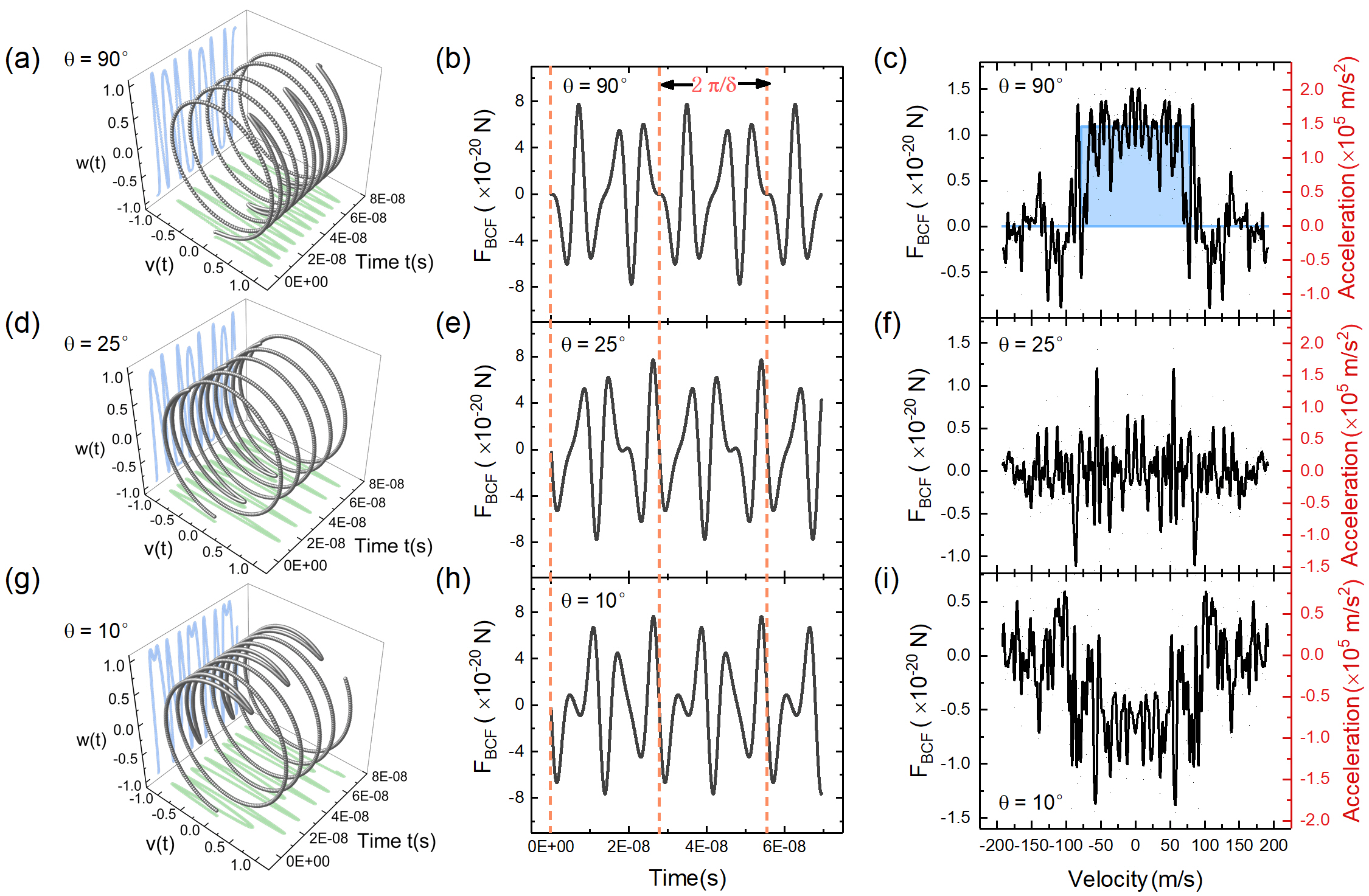}
\caption{\label{Fig:VBCF}
Left panels (a,d,g): Bloch vector evolution for initial phase values $\theta = 90^\circ, 25^\circ, 10^\circ$ (the system is initialized in the ground state and the evolution time of 80~ns is shown). Middle (b,e,h): instantaneous force $F_{\text{BCF}}$; orange dashed lines: $2\pi/\delta$ periods. Right (c,f,i): integrated net force vs. velocity (acceleration on right $y$-axis). Blue shaded region in (c): classical estimate (see \textit{appendix}), rectangle height = max acceleration, width = capture range.
}
\end{figure*}
The molecule-field interaction can be visualized and figure~\ref{Fig:VBCF} illustrates the Bloch vector evolution (represented on a Bloch cylinder~\cite{Galica2013PRA}), obtained by solving the time${-}$dependent optical Bloch equations (OBEs). 
The calculation framework is described in the \textit{appendix}. 
The results explicitly show the required phase reversal during the dynamics. Under optimal phase conditions, the Bloch vector first rotates counterclockwise by nearly $3\pi$, driven by the relative and initial phases of the optical fields. It then reverses direction and follows an almost identical return path. This four‑pulse periodicity originates from the phase alternation between successive pulses in the beat‑note train. Such controlled phase reversal ensures unidirectional net momentum transfer over each $6\pi$ optical cycle, producing a strong and sustained VBCF. The nearly constant amplitude of the Bloch vector confirms that quantum coherence is preserved throughout the light-molecule interaction. 
Figures~\ref{Fig:VBCF}(b) and \ref{Fig:VBCF}(c) show the velocity${-}$dependent instantaneous force (evaluated at the center of the molecular velocity distribution) and its time integral over the entire interaction window, respectively. Small spikes in the integrated force arise from weak multiphoton coupling. 
Because the VBCF results from integration over the entire interaction time, the initial phase plays a critical role by controlling the timing of phase reversal during the Bloch vector evolution, directly shaping the Bloch vector trajectory and the instantaneous force profile to yield net deceleration. This marks a qualitative distinction between the VBCF and conventional electronic‑transition BCF. 
Phase mismatch (Figures~\ref{Fig:VBCF}(d)--(f)) leads to destructive interference and a vanishing net force, whereas a reversed initial phase (Figures~\ref{Fig:VBCF}(g)--(i)) produces a net force in the opposite direction. 
In an experiment, the initial phase is determined by the arrival time of each molecule in the optical interaction region. Owing to the rapid field evolution and the finite velocity spread of the molecular beam, approximately 45\% of the molecules enter the light field with an initial phase favorable for deceleration, achieving an acceleration exceeding 80\% of the theoretical limit. This enables efficient VBCF deceleration.

\begin{figure}[bhtp]
\centering
\includegraphics[width=0.95\columnwidth]{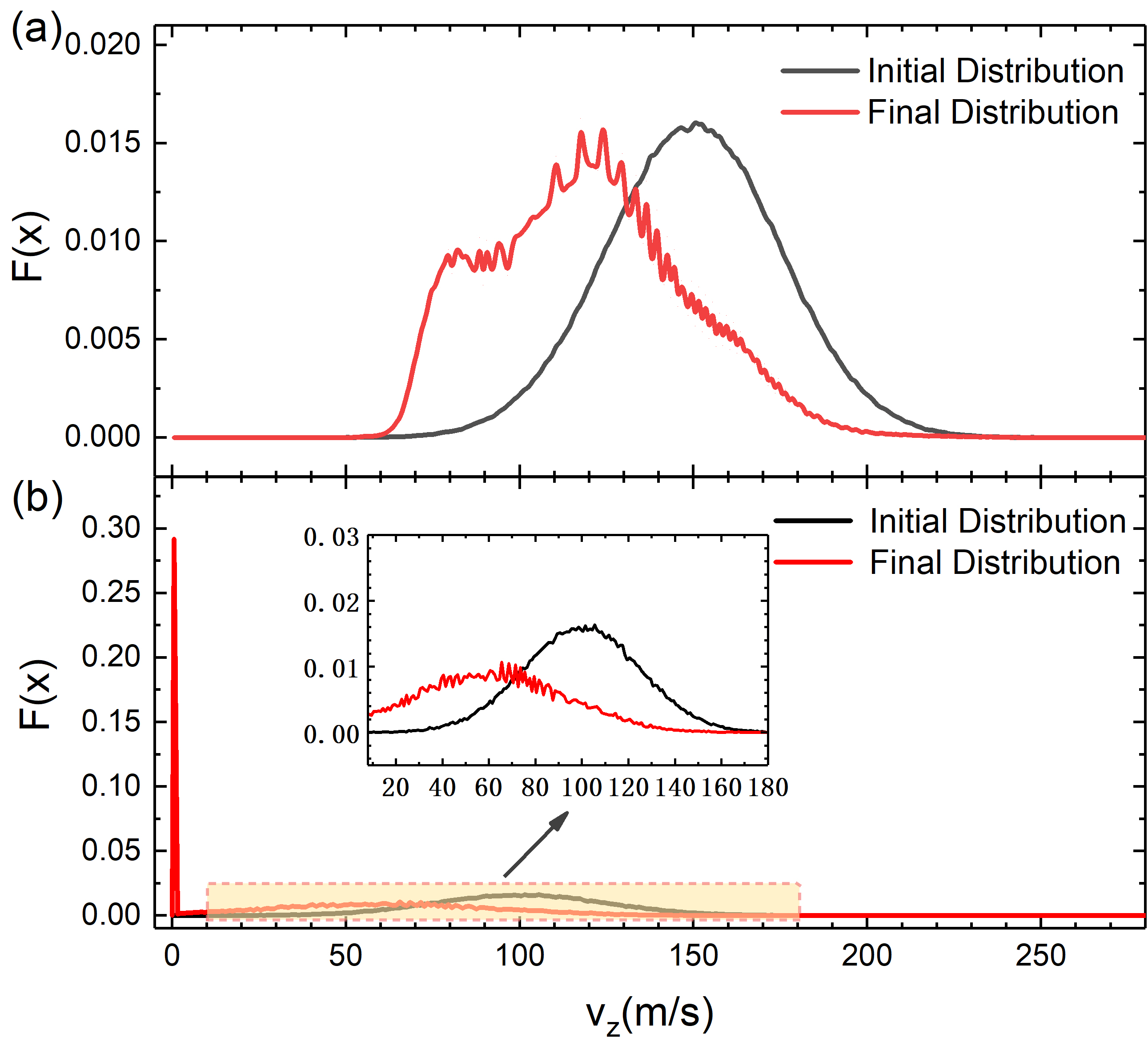}
\caption{\label{Fig:Velocity Simulation} 
Velocity distributions along the longitudinal direction (z direction) before (black) and after (red) deceleration via the VBCF. The simulation includes two different initial central velocities with the same $\pm$75~m/s range (three times the full width at half maximum of the distribution):(a) 150~m/s, (b) 100~m/s.
The oscillations superimposed on the decelerated distribution profile originate from the fluctuations in the numerically calculated VBCF. The inset in panel (b) provides a magnified view of the initial (black) and final (red) velocity distributions for the 100~m/s case.
}
\end{figure}

\section{Results}
We evaluated the deceleration performance using one-dimensional Monte Carlo simulations of the velocity distribution under optimized parameters. The model fully incorporates the coherent light-matter interaction underlying the VBCF while neglecting transverse beam losses, which do not affect the longitudinal dynamics (see \textit{appendix} for details). Key results are summarized in Figure~\ref{Fig:Velocity Simulation}. 
The VBCF remains nearly constant across the capture range but drops sharply at the lower boundary, where the Doppler shift moves the molecule outside the effective driving bandwidth. This sharp force cutoff pins the terminal velocity of slower molecules to this boundary, whereas faster molecules exit the deceleration zone with residual forward velocity before reaching it. Molecules with initial velocities within the capture range are decelerated over the interaction length. As shown in Figure~\ref{Fig:Velocity Simulation}(a), for the nominal beam velocity of 150~m/s, a substantial fraction of the ensemble undergoes efficient deceleration. For slower molecules, as shown in Figures~\ref{Fig:Velocity Simulation} (b), this velocity pinning produces pronounced compression of the distribution toward the terminal velocity --- a signature of collisionless cooling inherent to the VBCF mechanism. Our simulations further indicate that, by tailoring the capture bandwidth or implementing a cascaded multistage architecture, molecules can be decelerated to the meter-per-second regime (Figure~\ref{Fig:Velocity Simulation}(b)), directly enabling downstream trapping and further cooling. 

We further assessed the robustness of the VBCF through a comprehensive parameter scan across all critical control variables. To sustain the VBCF above 70\% of its peak value, the detuning must be stabilized to within 1.3~kHz, the relative phase between the counter-propagating beat trains to within $0.4^{\circ}$ (corresponding to a mirror displacement of 4.6~mm), and the optical intensity to within 1.0\%. A favorable initial phase --- set by the molecule's arrival time at the light field and absent in conventional electronic-state BCF, where the interaction is inherently dissipative --- is also required. Although arrival times are uniformly distributed over the optical cycle, our OBE analysis reveals that more than two-thirds of all possible initial phases yield effective deceleration, ensuring that a substantial fraction of molecules in a continuous beam encounter the field in a productive phase state.

Notably, for a given detuning $\delta$, numerical optimization can always identify a suitable parameter set, implying considerable flexibility in the VBCF configuration. In practice, we select $\delta$ values that correspond to experimentally accessible laser intensities. While these tolerances are more stringent than those of electronic-state BCF, they are well within current experimental capabilities. Full details of the OBE derivation, numerical framework, and input parameters are provided in the \textit{appendix}.

\section{Discussion}

\begin{table*}[htp]
\centering
\caption{\label{Tab:universal}
Calculated VBCF amplitude and acceleration of molecules and the laser power needed. The R(0) line of vibration fundamental band of each molecule is used. Transition parameters are taken from the HITRAN database~\cite{GORDON2026HITRAN2024}. 
}
\begin{ruledtabular}
\renewcommand{\arraystretch}{1.5}
\begin{tabular}{ccccccc}
 Molecules&Wavelength (nm)&Transition rate (s$^{-1}$)& Laser power$^a$ (W)  & $\delta/2\pi$ (MHz) & Capture range$^b$ (m/s) & Acceleration$^b$ (m/s$^2$)\\ \hline
 $^{12}$C$^{16}$O$_2$   & 4255.469   &  140.7    & 0.27 & 34.8 & $\pm$74.0 & $1.48\times 10^{5}$\\
 $^{12}$C$^{16}$O   & 4657.486  & 12.0     & 1.22 & 24.7 & $\pm$57.5& $1.51\times 10^{5}$\\
  $^{14}$N$_2$$^{16}$O   & 4495.215  & 70.7     & 0.50 & 36.4 & $\pm$81.9 & $1.47\times 10^{5}$\\
 H$_2$$^{16}$O   &2645.857   & 29.0     & 0.38 & 9.2 & $\pm$12.1 & $1.54\times 10^{5}$\\
 $^{12}$CH$_4$   &3301.690   & 25.4     & 0.28 & 10.3 & $\pm$17.0 & $1.55\times 10^{5}$\\
\end{tabular}
\end{ruledtabular}
\flushleft
$^a$ The laser power refers to the power of each beam and the focusing diameter is set to 0.5~mm.\\
$^b$ Estimated with the classical formulas shown in the \textit{appendix}.
\end{table*}

The non-dissipative approximation underlying the VBCF is valid when the laser-molecule interaction time is much shorter than the intrinsic relaxation time and the driving Rabi frequency far exceeds the relaxation rate. Under these conditions, spontaneous emission and population leakage are negligible. For fundamental vibrational transitions, the spontaneous emission rate is only $\sim 10$--$10^2$~s$^{-1}$, while the VBCF cycling rate reaches $\sim 10^7$--$10^8$~s$^{-1}$; thus, over the full $\sim 10^4$ cycles, total population leakage remains negligible. Even when rotational or hyperfine structure is included, the dynamics remain accurately described by an effective two-level model. For $^{13}$CO$_2$, the ratio of vibrational relaxation rate to Rabi frequency is $10^{-5}$ --- orders of magnitude smaller than the per-cycle leakage in electronic-transition deceleration (typically 0.01--1). Hyperfine splitting (e.g., 50~kHz for the $J=1$ excited state) induces only a $\sim$14-cycle phase slip over the entire interaction, yielding no appreciable effect on deceleration. This inherent robustness against state complexity is a defining feature of the VBCF.

The VBCF scheme exhibits exceptional universality: it applies to any molecular species with an allowed vibrational transition, including both polar and weakly polar molecules. Table~\ref{Tab:universal} lists the calculated VBCF parameters for several representative molecules --- CO$_2$, CO, N$_2$O, H$_2$O, and CH$_4$ --- whose fundamental vibrational transitions fall within the 2--5~$\mu$m wavelength range. For these species, conventional mid-infrared laser beams with a power of 1.5~W and a focusing diameter of 0.5~mm meet the requirements for VBCF. For molecules with smaller transition dipole moments, efficient deceleration can still be achieved by scaling up the driving laser intensity.

Compared to existing molecular deceleration techniques, the VBCF offers a distinct combination of advantages and trade-offs. Stark and Zeeman decelerators deliver comparable forces ($10^4$--$10^5$~m/s$^2$), but they require complex multi-electrode arrays and are restricted to polar or paramagnetic species~\cite{Bethlem1999PRL, Bethlem2000PRL, Bethlem2000Nature, doyle1995, hogan2008magnetic}. Buffer-gas cooling produces bright, versatile beams yet typically cannot cool below $\sim$100~m/s~\cite{Hutzler2012CR}. Direct laser cooling via electronic transitions achieves sub-mK temperatures but only in a handful of molecules with quasi-closed cycling transitions~\cite{Horchani2025QS, Vilas2022Nature, Collopy2018PRL, Shuman2010Nature, Lim2018PRL, Carson2022NJP, Augenbraun2020NJP, Kozyryev2017PRL, Mitra2020Science, Stuhl2008PRL, Tsikata2010NJP, Hummon2011PRL}. The VBCF bridges these gaps: it generates a coherent force of $\sim 10^5$~m/s$^2$ from a single millimeter-scale optical stage and is applicable to any molecule with a moderate vibrational transition. Its main limitations are stringent phase-control requirements and, currently, confinement to one-dimensional longitudinal deceleration. With cascaded multi-stage architectures and transverse confinement, the VBCF could provide a broadly applicable, all-optical pathway to cold molecular samples, complementing existing methods.

Laser deceleration via vibrational transitions has long been hindered by the lack of high-power mid-infrared sources capable of precise, deterministic frequency and phase control. Recent advances in precision mid-infrared optical parametric oscillator (OPO) technology~\cite{ZhangOL2020, zhangOE2020, Tan2024CJCP-OPO} now meet the stringent demands of the VBCF scheme. We have demonstrated phase- and frequency-locked mid-infrared OPOs with sub-kHz frequency stability, narrow linewidths, watt-level output power, and fast feedback for active phase and intensity control. These capabilities directly fulfill all technical requirements identified in our robustness analysis, providing a clear path to experimental realization. The phase-stabilized optical architecture is illustrated in the \textit{appendix}.

\section{Conclusion}
In summary, we demonstrate a molecular deceleration scheme based on the vibrational transition‑mediated bichromatic force (VBCF). Numerical analysis shows that this scheme efficiently decelerates molecules and simultaneously produces velocity compression characteristic of collisionless cooling. Using a single millimeter‑scale optical stage, the VBCF generates a coherent force of \(\sim1.45\times10^5\ \text{m/s}^2\), and is applicable to any molecule with a strong fundamental vibrational transition. The main challenges are stringent phase control and, at present, confinement to one‑dimensional longitudinal deceleration. Our VBCF framework unlocks deceleration and coherent manipulation of a wide range of molecular species inaccessible to conventional techniques, enabling transformative advances in cold molecular physics and precision metrology. Integration with state‑of‑the‑art coherent control tools will further extend molecular coherence manipulation, establishing a versatile platform for quantum computation, quantum simulation, and tests of fundamental physics.


\begin{acknowledgments}
The manuscript's language and readability were partially refined with the assistance of AI. The authors thank W. Jiang, T. Xia, and Y.R. Sun for the discussion. This work was jointly supported by the Chinese Academy of Science (Grant No.YSBR-055), the Innovation Program for Quantum Science and Technology (Grant No. 2021ZD0303102), the National Natural Science Foundation of China (Grant Nos. 22241302, 12393825, 12504569), and the China Postdoctoral Science Foundation (Grant No. 2024M753081).
\end{acknowledgments}

C.-F.C. and S.-M.H. designed the research; M.-Y.Y. and C.-F.C. developed the theory; all authors discussed the results; and M.-Y.Y., C.-F.C, and S.-M.H. wrote the paper. 

Data supporting the findings of this article are openly available.

\clearpage
\appendix

\section{\label{Classic expression}Classic expression of VBCF}

The commonly used theoretical descriptions of BCF in atoms and related quantum systems include the $\pi$-pulse model and the dressed-state theory~\cite{voitsekhovich1989observation, Grimm1997PRL}. These theoretical frameworks are also applicable to the vibrational bichromatic force (VBCF). Fig.1a of the main text depicts the VBCF configuration: two lasers with slightly different frequencies propagate in opposite directions, equivalent to four independent monochromatic plane waves in the laboratory frame. For this symmetric detuned bichromatic field, the total electric field is the superposition of two beat wave trains, which can be expressed as:

\begin{equation}
\begin{aligned}
E(z,t)=2E_0~cos[w(t-z/c)]cos[\delta(t-z/c)+\phi/4]\\
+2E_0~cos[w(t+z/c)]cos[\delta(t+z/c)-\phi/4]\\
\end{aligned}
\label{EquS_field}
\end{equation}\\

The offset terms $\pm\phi/4$ in this equation directly determine the delay between the two oppositely propagating beat trains.
By precisely controlling the relative phase to $\phi=\pi/2$, the direction of momentum transfer in the absorption and stimulated emission processes can be regulated, which is a core condition for the implementation of VBCF.

In the $\pi$-pulse model, when each beat with a duration of $\pi/\delta$ is tuned to be a complete $\pi$-pulse, complete population transfer--either from the ground state to the excited state or vice versa--is achieved. These momentum transfer and accumulation processes cause the BCF deceleration. The condition for forming a $\pi$-pulse within one beat period is expressed as:

\begin{equation}
\int_{-\pi/{2\delta}}^{\pi/{2\delta}}2\Omega_0 cos(\delta t)dt=\pi
\end{equation}\\

This integral yields the constraint between the Rabi frequency and the detuning:~$\Omega_0=\pi\delta/4$. 

By decomposing the interaction into individual beats, the $\pi$-pulse model offers a qualitative understanding of the momentum transfer process, but it neglects the continuous nature of the quantum evolution. 

A more rigorous approach is the dressed-state model proposed by Grimm et al.~\cite{Grimm1994}. Combined with Floquet analysis, this model describes the BCF deceleration effect as a potential-energy climbing process with level spacing $\hbar\delta$~\cite{Metcalf2004}. By numerically solving the eigenstates of the Hamiltonian, the dressed-state theory not only confirms the optimal relative phase $\phi=\pi/2$ but also yields a more accurate prediction for the optimal Rabi frequency: $\Omega_0=\sqrt{3/2}~\delta$. 

Owing to the periodic amplitude modulation of the superposed beat trains, the potential-energy trajectory comprises a 3/4 climbing phase (molecular deceleration) and a 1/4 descending phase (molecular acceleration)~\cite{Metcalf2004}. The resulting bichromatic force is therefore expressed as:

\begin{equation}
F_{BCF}=\frac{\Delta E}{\Delta x}=\frac{2\hbar\delta}{\lambda}=\frac{\hbar k \delta}{\pi}
\label{EquA_Force}
\end{equation}
Additionally, the dressed-state theory predicts the velocity capture range, 
\begin{equation}
\Delta v_c=\delta/k
\label{EquS_Range}
\end{equation}

For conventional BCF schemes that rely on electronic transitions, spontaneous emission is unavoidable and reduces the overall deceleration efficiency. 
By contrast, for the VBCF based on vibrational transitions, the condition $\gamma\ll\Omega_0$ ensures the continuity and integrity of the bichromatic force interaction, allowing the experimental deceleration to closely follow the classical theoretical predictions.


\section{\label{Bloch evolution}Numerical solution of VBCF}

Within the rotating-wave approximation (RWA), the optical Bloch equations (OBEs) provide a precise numerical description of the internal state evolution of a two-level system. Derived from the quantum Liouville equation including dissipation, this method~\cite{Grimm1997PRL} fully incorporates Rabi cycling and spontaneous emission effects. 

\begin{equation}
\begin{aligned}
\dot{u}&=-\Delta_{12}v-\text{Im}[\Omega_{12}]w-\frac{\gamma_{12}}{2}u\\
\dot{v}&=\Delta_{12}u+\text{Re}[\Omega_{12}]w-\frac{\gamma_{12}}{2}v\\
\dot{w}&=\text{Im}[\Omega_{12}]u-\text{Re}[\Omega_{12}]v-\gamma_{12}(1+w)\\
\end{aligned}
\end{equation}\\

Here, $\Delta_{12}$ denotes the detuning between the central bichromatic frequency and the resonant transition frequency. The Rabi frequency associated with the electric field is given by:  

\begin{equation}
\begin{split}  
    \Omega_{12}&=\frac{\mu_{12} E(z,t)}{\hbar}\\
    &=\frac{4\mu_{12} E_0}{\hbar}[\cos(kz)\cos(\delta t+\theta)\cos(\phi/4)
    \\ &\qquad \qquad+i \sin(kz)\sin(\delta t+\theta)\sin(\phi/4)]\\
    &=4\Omega_{0} [ \cos(kz)\cos(\delta t+\theta)\cos(\phi/4)
    \\ &\qquad \qquad+i \sin(kz)\sin(\delta t+\theta)\sin(\phi/4)],
\end{split}
\end{equation}
where $\mu_e$ is the transition dipole matrix element, and the additional initial phase $\theta$ represents the envelope phase of the electric field at the onset of the interaction. To illustrate the effect of $\theta$, we simulate the Bloch sphere evolution for three representative initial phase values as shown in Fig.2. The components of the Bloch vector $(u, v, w)$ correspond to density matrix elements, defined as: 

\begin{equation}
\begin{aligned}
    u\equiv&~\rho_{12}+\rho_{21}\equiv2\text{Re}[\widetilde{\rho}_{12}]\\
    v\equiv&~i(\rho_{12}-\rho_{21})\equiv2\text{Im}[\widetilde{\rho}_{12}]\\
    w\equiv&~\rho_{22}-\rho_{11}\\
\end{aligned}
\end{equation}

The $\nu(t)$ component describes the energy exchange rate between the molecule and the optical field.
Together with the $w(t)$ component, which reflects the degree of population inversion, they form the ``Bloch cylinder'' diagram, providing an intuitive visualization of the bichromatic force evolution. 

By solving the OBEs over small time steps, the instantaneous force can be evaluated via Ehrenfest's theorem~\cite{Metcalf2003}:

\begin{equation}
F(z,t)=\hbar[u(t)\Delta \text{Re}[\Omega_{12}(z,t)]-v(t) \text{Im}[\Omega_{12}(z,t)]]
\end{equation}

The first term in this equation corresponds to a high-frequency component that causes rapid oscillations in the time-domain force curve. However, the influence of this component cancels out upon time integration, leaving a non-zero average net force determined by the second term. 
With this expression, we simulate the time-domain evolution of the bichromatic force.

\section{\label{Robustness}Robustness of VBCF}

\begin{figure*}[htbp]
\centering
\includegraphics[width=0.95\textwidth]{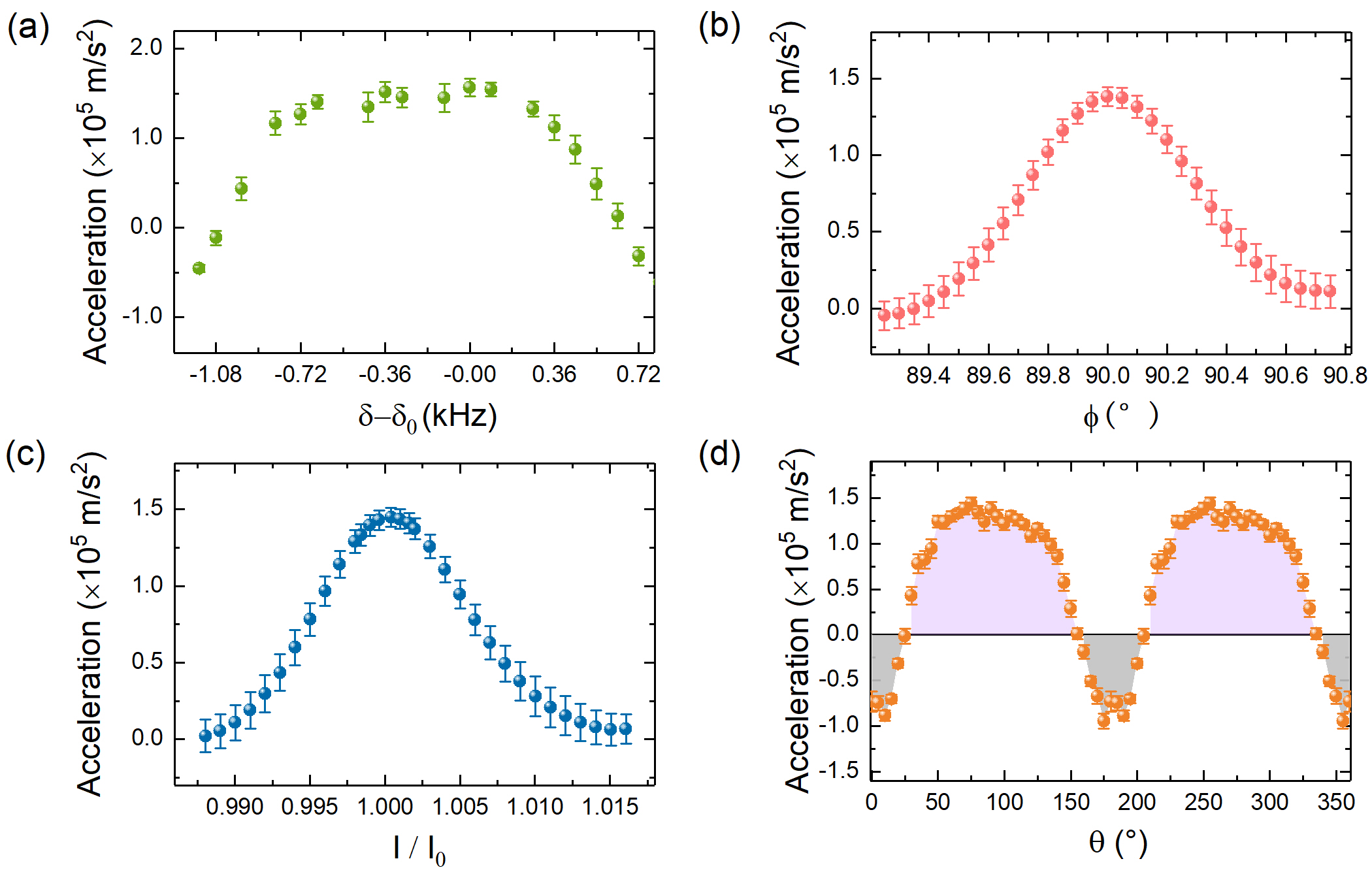}
\caption{\label{Fig:Parameters}
Parameter robustness analysis of VBCF. 
(a) Acceleration as a function of detuning frequency deviation $\delta-\delta_0$. 
(b) Acceleration as a function of relative phase between counter-propagating beat trains. 
(c) Acceleration as a function of normalized optical intensity $I/I_0$.  
(d) Acceleration as a function of the initial phase of the beat trains. The purple and gray shadows indicate the probability of molecules in the region of plus and minus acceleration towards the molecular flying direction, respectively.
Note that the error bar of each calculation point indicates the standard deviation of the averaged acceleration in the capture velocity ranging from -75~m/s to 75~m/s. 
}
\end{figure*}

To quantitatively assess the robustness of the VBCF deceleration scheme, we perform a comprehensive numerical analysis of its sensitivity across the full set of critical control parameters. 
We solve the time-dependent optical Bloch equations (OBEs) incorporating vibrational level structure and state-dependent relaxation rates for the $^{13}$CO$_2$ fundamental transition, to compute VBCF magnitude and velocity dependence across parameter space. And these complete parameter sensitivity results are summarized in Fig.~\ref{Fig:Parameters}.
The detuning $\delta$ scales approximately linearly with the optical force magnitude. 
Excessively large $\delta$ limits coherent cycling accessibility and renders precise phase control experimentally intractable. 
Fig.~\ref{Fig:Parameters}a shows that a detuning control accuracy of 1.3~kHz is required to preserve VBCF performance.
The relative phase between counterpropagating beat-note trains is critical for maintaining continuous coherent absorption–emission cycling. Deviations disrupt Bloch vector phase reversal, reduce coherent transfer cycles, and degrade peak force (Fig.~\ref{Fig:Parameters}b). In our scheme, the beating spatial period is $c\pi/\delta$=4.16~m. A 30\% reduction in peak acceleration corresponds to a relative phase deviation of $\pm$0.2$^\circ$, equivalent to a path-length deviation of 4.6 mm--well within standard translation-stage and active phase-locking precision.
VBCF magnitude and achievable acceleration depend sensitively on coherent radiation cycles and Rabi frequency, and thus on incident optical intensity (Fig.~\ref{Fig:Parameters}c). For fixed detuning, an optimal intensity $I_0$ exists uniquely. For $^{13}$CO$_2$, we calculate $I_0$ = 302.8~W/cm$^2$ per laser beam, which corresponds to a 297~mW laser focus in a beam diameter of 0.5~mm. A mere $\pm$0.5\% deviation in optical intensity induces about a 30\% drop in acceleration.
The initial-phase dependence is summarized in Fig.~\ref{Fig:Parameters}d. Within one beat period, the effective force dominates over two-thirds of the cycle (shaded region). Critically, this relative scaling remains valid in the strong-driving regime where the Rabi frequency greatly exceeds the vibrational-state relaxation rate. And that ensures a large fraction of molecules naturally enters the interaction region with a favorable phase for efficient deceleration, without ultrafast gating or molecular beam bunching.

\section{\label{Monte Carlo}Simulation of velocity distribution}

\begin{figure}[htbp]
\centering
\includegraphics[width=0.95\linewidth]{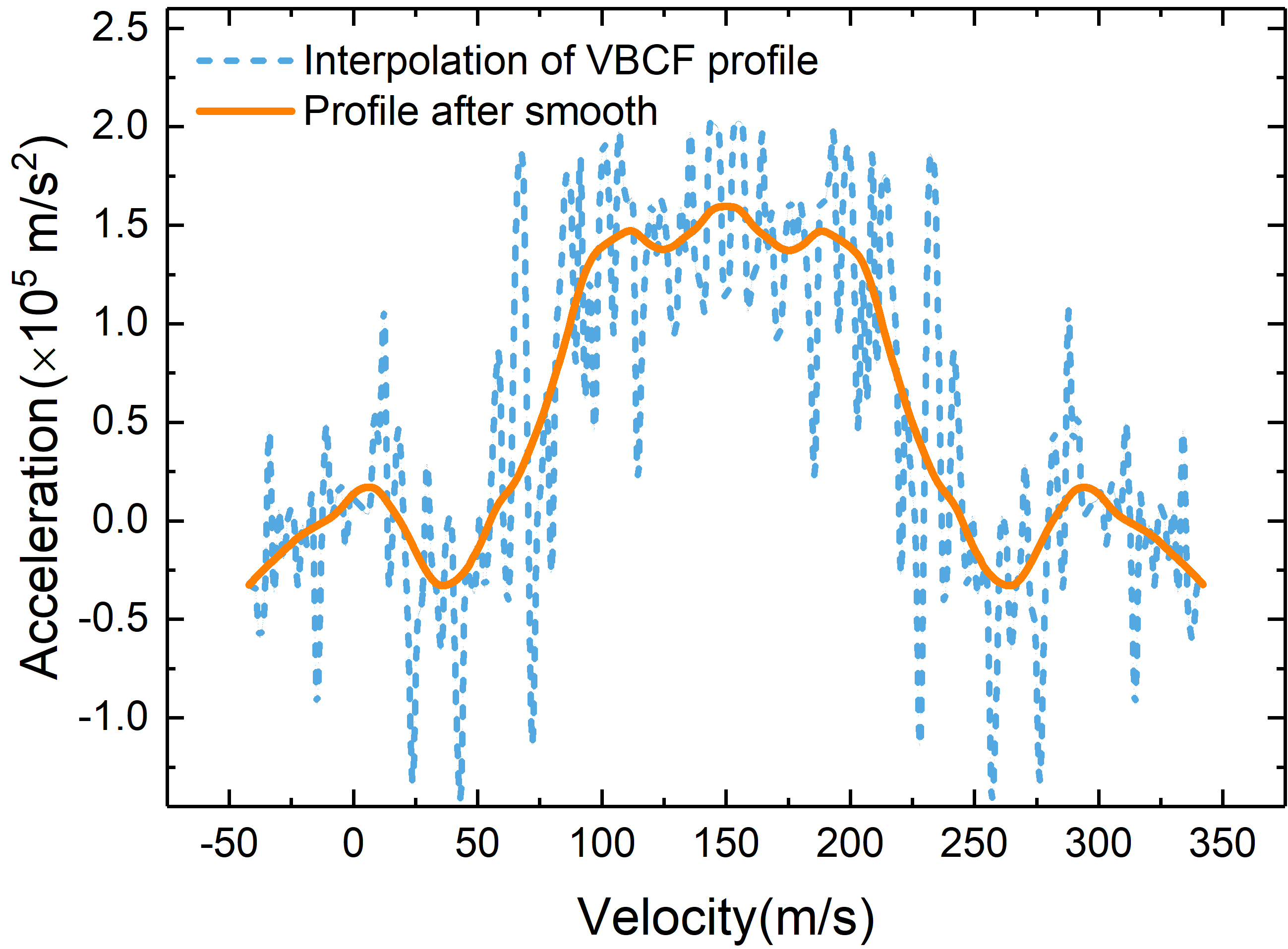}
\caption{\label{Fig:Smooth}
The VBCF profile after interpolation and smoothing}
\end{figure}

The simulated forward velocity distributions of the molecular beam shown in Fig. 2 illustrate the deceleration effect of VBCF. This one-dimensional distribution is obtained through the following procedure: \\
1. A sufficient number (around 5 million molecules) is set to satisfy a velocity distribution of $150\pm75~m/s$~($3\sigma$);\\ 
2. The initial velocity distribution is divided into sufficiently fine bins ($\Delta v_f=0.5~m/s$), with molecules within each bin assumed to experience the same deceleration;\\
3. The numerically calculated BCF profiles are interpolated and smoothed, as shown in Fig~\ref{Fig:Smooth}, providing the acceleration value for each velocity bin;\\ 
4. The deceleration process is discretized into subintervals of length $\Delta t=20~\pi/\delta$ in the time domain. After each deceleration subinterval, the acceleration corresponding to each velocity bin is updated by returning to Step 3;\\
5. Repeat Step 4 until the molecules fly out of the interaction region after undergoing multiple deceleration subintervals. 
The final decelerated molecular beam velocity distribution is obtained by summing over all velocity bins, yielding the red curve in Fig. 2. 


\section{\label{Lasers} Laser system for experimental realization}
\begin{figure}[htbp]
\centering
\includegraphics[width=0.95\linewidth]{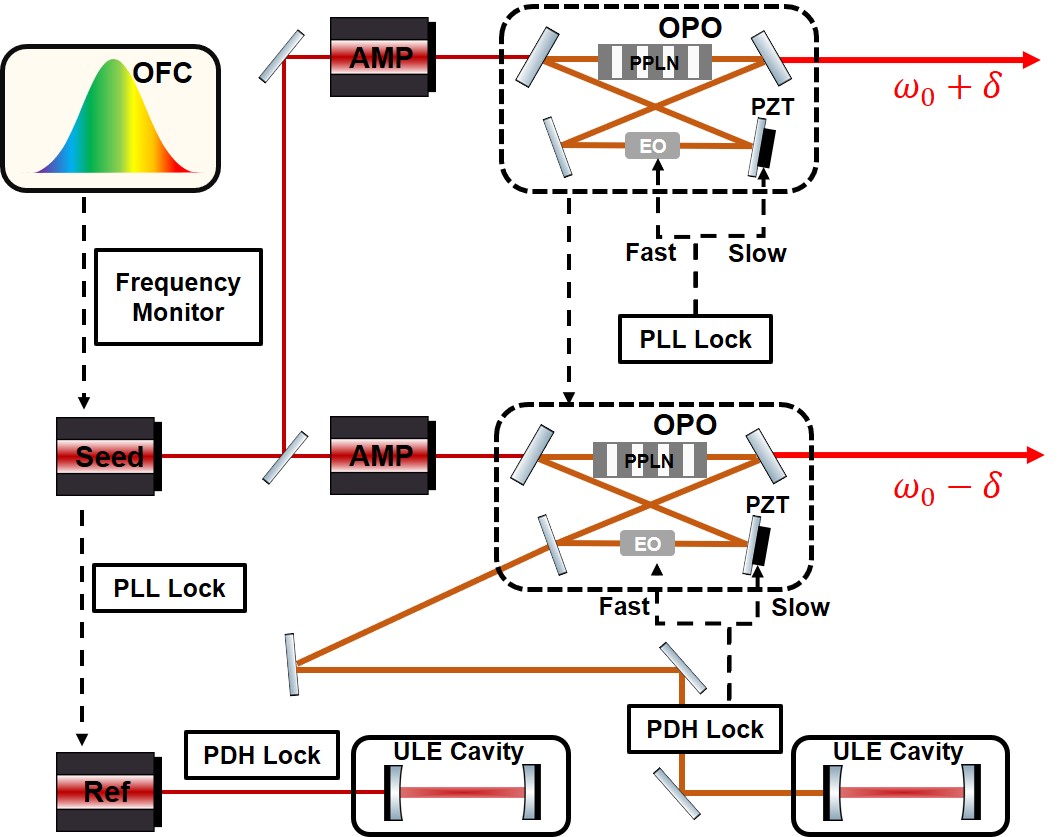}
\caption{\label{Fig:Optics}
The schematic diagram of the optical design for each beat note.}
\end{figure}
The VBCF scheme imposes stringent requirements on the driving light fields: phase coherence between counter-propagating beat notes, ultra-stable absolute optical frequency, and precisely tunable bichromatic detuning $\delta$. We realize this via a fully phase-stabilized mid-infrared optical architecture, with the full schematic presented in Fig.~\ref{Fig:Optics}. 
The system is seeded by a narrow-linewidth 1064 nm Nd:YAG laser (instantaneous linewidth $\leq$ 1 kHz), which is split into two equal-power channels. Each channel is amplified to 50~W via a high-power fiber amplifier, and used to pump a home-built singly resonant optical parametric oscillator (OPO) for the generation of mid-infrared light resonant with the molecular vibrational transition. Each OPO integrates a dual-bandwidth feedback system for frequency and phase control: an intra-cavity electro-optic (EO) crystal with a 5~MHz bandwidth for high-speed phase corrections, and a piezoelectric transducer (PZT) with a 10~kHz bandwidth attached to a cavity mirror for slow, long-term cavity length stabilization. 
To establish an absolute frequency reference, the signal beam of the master OPO is stabilized to a high-finesse ultra-low-expansion (ULE) optical cavity using the Pound-Drever-Hall (PDH) technique, suppressing both fast frequency noise and slow thermal drift. The signal beam of the slave OPO is then phase-locked to the master OPO via a digital phase-locked loop (PLL). The derived error signals are split into high- and low-frequency components, routed to the slave OPO's EO crystal and PZT actuator respectively, to maintain tight phase lock over the full interaction time of the deceleration experiment. The radio-frequency reference for the PLL is set to $2\delta$, which precisely defines the $\pm\delta$ frequency detuning of the bichromatic beat notes relative to the molecular vibrational resonance, as required for VBCF. 
To eliminate long-term drift of the 1064~nm pump system, an independent narrow-linewidth 1064~nm reference laser is stabilized to a second ULE optical cavity. The 1064 nm seed laser is phase-locked to this reference laser, with its absolute optical frequency continuously monitored via a femtosecond optical frequency comb. This enables deterministic, in-situ tuning of the absolute mid-infrared frequency to match the molecular transition, with sub-kHz long-term frequency stability. 
This fully stabilized optical architecture delivers a pair of phase-coherent mid-infrared laser beams with precisely controlled frequency detuning, ultra-low phase noise, and well-characterized absolute optical frequency, fully satisfying the requirements for coherent, sustained VBCF-driven molecular deceleration.

\end{CJK*}

\bibliography{BCF26.bib}


\end{document}